# Sensitivity of quantitative diffusion MRI tractography and microstructure to anisotropic spatial sampling

Authors: Elyssa M. McMaster[1,*], Nancy R. Newlin[2], Chloe Cho[3], Gaurav Rudravaram[1], Adam M. Saunders[1], Aravind R. Krishnan[1], Lucas W. Remedios[2], Michael E. Kim[2], Hanliang Xu[2], Kurt G. Schilling[4,5], François Rheault[6], Laurie E. Cutting[1,5,7,8], and Bennett A. Landman[1,2,3,4,5]

[1] Department of Electrical and Computer Engineering, Vanderbilt University, Nashville, TN USA
[2] Department of Computer Science, Vanderbilt University, Nashville, TN USA
[3] Department of Biomedical Engineering, Vanderbilt University, Nashville, TN USA
[4] Vanderbilt University Institute of Imaging Science, Vanderbilt University Medical Center, Nashville, TN USA
[5] Department of Radiology and Radiological Sciences, Vanderbilt University Medical Center, Nashville, TN USA
[6] Department of Computer Science, Université de Sherbrooke, Sherbrooke, QC, Canada
[7] Vanderbilt Kennedy Center, Vanderbilt University, Nashville, TN USA
[8] Peabody College of Education, Vanderbilt University, Nashville, TN USA

Word Count: 3,214
*Corresponding author

Abstract

Purpose: Diffusion weighted MRI (dMRI) and its models of neural structure provide insight into human brain organization and variations in white matter. A recent study by McMaster, et al. showed that complex graph measures of the connectome, the graphical representation of a tractogram, vary with spatial sampling changes, but biases introduced by anisotropic voxels in the process have not been well characterized. This study uses microstructural measures (fractional anisotropy and mean diffusivity) and white matter bundle properties (bundle volume, length, and surface area) to further understand the effect of anisotropic voxels on microstructure and tractography.

Methods: The statistical significance of the selected measures derived from dMRI data were assessed by comparing three white matter bundles at different spatial resolutions with 44 subjects from the Human Connectome Project – Young Adult dataset scan/rescan data using the Wilcoxon Signed-Rank test. The original isotropic resolution (1.25 mm isotropic) was explored with 6 anisotropic resolutions with 0.25 mm incremental steps in the $z$ dimension. Then, all generated resolutions were upsampled to 1.25 mm isotropic and 1 mm isotropic.

Results: There were statistically significant differences between at least one microstructural and one bundle measure at every resolution ($p \leq 0.05$, corrected for multiple comparisons). Cohen's $d$ coefficient evaluated the effect size of anisotropic voxels on microstructure and tractography.

Conclusion: Fractional anisotropy and mean diffusivity cannot be recovered with basic up-sampling from low quality data with gold-standard data. However, the bundle measures from tractogram become more repeatable when voxels are resampled to 1 mm isotropic.

1. Introduction

Diffusion weighted MRI (dMRI) evaluates the degree of random micron-scale movement of water within tissue [1]. Several established methods use the data collected from dMRI to approximate the random thermal motion of water molecules within the brain and model white matter tracts, including microstructural settings such as diffusion tensor imaging (DTI) and fiber orientation distribution (FOD) functions, and in white matter macrostructure with tractography [2], [3]. These approaches have been used to gain fundamental knowledge about the brain and its connections to better understand the human brain map as well as white matter diseases and disorders [4], [5], [6]. Despite their widespread study and implementation, these approximations often suffer from a lack of reproducibility and repeatability due to variations in scanner protocol between sites and studies [7]. One source of this variation is the acquisition's spatial sampling, specifically between isotropic and anisotropic sampling [8].

The effect of spatial sampling on image-based brain map approximations can be observed in the microstructure [9]. Microstructural metrics from DTI are computed on a voxel-wise basis. The diffusion tensor fits a 3x3 matrix that describes the principal directions of diffusion [10]. A larger voxel means that a tensor is responsible for representing a larger amount of the brain's area, and the larger scale can lead to less sensitive representations of diffusion. If the voxel dimensions are anisotropic, it can allow bias along the larger directions that will be reflected in the tensor images and tensor metrics and quantitative maps, including fractional anisotropy (FA) and mean diffusivity (MD).

Tractography, a process that uses diffusion imaging to generate maps of the brain's white matter fibers, starts with either the DTI or FOD maps of an image, which means that the microstructure is considered at a brain-wide scale. The tractography algorithm generates streamlines to show the highest direction of diffusion that may correspond with the location of white matter fibers [11], [12]. Without a reliable microstructural foundation, the tractography algorithm may be more susceptible to false-positive or spurious streamline generation that does not interpret the white matter tracts in a meaningful or representative way.  DTI and FODs each are both voxel-wise operations, meaning that every voxel is represented with either a diffusion tensor or FOD function glyph [13], [12], [14]. Larger voxels can introduce multiple tissue types or fibers into the same voxel space which makes the tensor or glyph more susceptible to biases from other tissues that fail to represent the true anatomy [15]. It is possible that anisotropic voxels may introduce biases into the computations of the tensors or glyphs that make downstream approximations inaccurate for representing anatomy in microstructure and macrostructure (Figure 1).

We can understand how the downstream models of white matter from diffusion MRI, such as DTI, FODs, and tractography, rely on the spatial sampling of an image with this intuitive explanation, but these effects have not been empirically studied. McMaster, et al. executed a systematic study on the impact of isotropic voxel size on the complex graph measures of the tractogram's graphical representation, a connectome, to analyze and address variance in the representation of white matter. Most of the complex graph measure calculations in whole-brain tractography became more reproducible with voxels reshaped to 1 mm isotropic, even in down sampled, lower quality data. The study suggests that spurious and false-positive streamlines in tractography result from voxel-wise operations assigned by the algorithm, not an information loss problem from lower quality data. To test the assertion that voxel size can cause variability in tractography, data taken from the same subject's scan and resampled to acquisition resolutions of major national studies were used to generate tractography, as well as the same image resampled to 1 mm isotropic. This experiment showed significant variation in the approximation of white matter tracts and complex graph measures within the connectome. Though data appears more reproducible when resampled from both isotropic and anisotropic voxels in a qualitative assessment, the specific biases introduced by anisotropic voxels have not yet been articulated [8]. It is essential to understand these biases, as many major studies use anisotropic voxels as part of their acquisition protocol in addition to widespread use of anisotropic voxels in clinical settings [8], [16], [17].

Other studies have examined the sensitivity of DTI and tractography metrics to scanner acquisition parameters, but none have articulated the specific biases introduced by anisotropic voxels in DTI and tractography in modern dMRI. A 2001 study observed the decrease in partial volume effects (PVE) in a DTI study with anisotropic voxels, with voxel sizes $3.75 \times 3.75 \times 7$ $mm^3$ and $1.88 \times 1.88 \times 5$ $mm^3$ [18]. This study contributed an important principle for future DTI study, but scanner technology has improved enough to explore this principle in higher resolutions, even in clinical data. The problem of PVE due to heterogeneous tissue types and fiber proximity has been articulated [19] and investigated [15], [20], but these studies have demonstrated these issues with isotropic voxels. Dyrby, et al. have reported successful reconstruction of anatomical details in an *ex vivo* monkey brain with interpolation of an isotropic acquisition, but contemporary scanner technology of 2014 could not capture images of *in vivo* human brain to reasonably recapture anatomical detail to perform a voxel-wise microstructure analysis task [21]. Neher et. al demonstrated that anisotropic voxels with dramatically elongated third dimensions introduce biases in phantom tractography datasets and found that the effect of the anisotropic voxels could be mitigated with up sampling to an isotropic resolution, but did not use *in vivo* data across a population or multiple tracts that may have a range of susceptibility to the direction of anisotropy [22]. McMaster, et. al showed sensitivity of the connectome's complex graph measures to spatial sampling changes but did not use microstructural measures nor measure the effect sizes of anisotropic voxels compared to isotropic [8].

This work diverges from previous work with the quantification of different approximations of white matter not previously studied by this strategy (fractional anisotropy, mean diffusivity, and bundle metrics) [8]. Though complex graph measures can help as a metric of reproducibility, they neglect the quantitative microstructural maps that have their own sensitivities to voxel sizes because tractography is often generated from FODFs instead of DTI. Repeatability in complex graph measures do not necessarily describe changes in bundle surface area, volume, and average streamline length that may be useful in reproducibility studies. In this study, we test the consistency of this approach across downstream approximations of white matter in microstructure and macrostructure beyond complex graph measures. We propose a study to measure the bias of anisotropic voxels on representations of the brain's microstructure and tractography bundles and then measure the efficacy of resampling their respective metrics.

## 2. Methods

We hypothesize that the spatial sampling of a diffusion MRI image will introduce statistically significant effects in the DTI metrics and bundle metrics due to the interrelated nature of spatial resolution and tractography shown by McMaster, et. al in 2024 [8]. We have established that up sampling an image to have more voxels, and therefore more tensors or FODs, can mitigate partial volume effects and minimize heterogeneity of tissue types in a single voxel space [8], [15], [23]. This study implements more DTI and bundle-related approaches to evaluate the effect of anisotropic voxels on evaluations of diffusion data. We implement the proposed strategy to resample voxels to 1 mm isotropic for DTI and tractography tasks and show the reduced variability in these metrics in addition to the more repeatable complex graph measures as shown previously in our isotropic dataset (Figure 2). We resample the original HCP data to anisotropic voxels in 0.25 mm steps in the third dimension. We evaluate the effect size of the anisotropy with the Cohen's $d$ coefficient between the distributions of each metric across run 1 of the scan-rescan data, and we use the effect size between the same resolution's run 1 and run 2 as a comparison baseline.

2.1 Data

The diffusion image dataset consisted of 44 healthy subjects (32 female) of the scan-rescan data from the Human Connectome Project – Young Adult (HCP-YA) dataset. HCP used their own diffusion imaging paradigm with b-values 1000, 2000, and 3000 s/mm$^2$ with a 3T WU-Minn-Ox HCP resolution of 1.25 $x$ 1.25 $x$ 1.25 mm$^3$. HCP used their minimal preprocessing pipeline on the diffusion images to correct eddy current distortions from the diffusion gradient. Average age among subjects is 30.36 years with standard deviation ± 3.34, as the HCP-YA dataset only includes ages 22-35.

2.2 Preprocessing

Though the diffusion data from HCP is already preprocessed, we also preprocessed with PreQual [24] for efficient quality assurance and diffusion tensor analysis and visualization. Once the initial HCP-YA dataset had been preprocessed with PreQual and Freesurfer, we used *mrgrid* from MRTrix3 to down sample the voxel sizes of each of the 44 scan/rescan subjects' images [12], [25], [26]. Five additional datasets were generated by increasing the voxel sizes of the images in 0.25 mm increments (1.25 mm isotropic to 1.25 mm $x$ 1.25 mm $x$ 2.75 mm anisotropic). The McMaster et al. study showed that the complex graph measures of voxels longer than 3 mm in any direction became unstable and unsuitable for practical study. In the case of down-sampling, *mrgrid* uses Gaussian smoothing as its default interpolation method [26]. After the down sampled images were preprocessed, we simulated a study with preprocessed low-resolution data by reshaping the voxels of the lower quality images to 1.25 mm isotropic and resampled all low-resolution data and original HCP data to 1 mm isotropic. When the image is up sampled, we select the *mrgrid* default interpolation method, cubic interpolation [26].

2.3 DTI Metrics

We fit a tensor model to our preprocessed data with MRTrix3's *dwi2tensor* [27]. We generate fractional anisotropy (FA) and mean diffusivity (MD) quantitative maps with *tensor2metric* [28]. We threshold b-values to be <=1500 for the quantitative maps. We collect mean of the FA and MD measurements with scilpy [29].

2.4 Tractography

We generate tractography bundles using TractSeg [30], [31] with the images, b-values and b-vectors preprocessed by PreQual [24]. We use scilpy to evaluate individual bundle metrics for bundle volume, bundle length, and bundle surface area and for DTI metrics over all TractSeg bundles.

2.5 Statistical Analysis

We perform a Wilcoxon Signed-Rank test [32] to compare the original resolution images and the down sampled images. The null hypothesis asserts that the median differences between paired observations of changes in spatial resolution has no effect on microstructural or macrostructural metrics of bundles, while the alternative hypothesis asserts that the differences between paired observations is not zero. We reject the null hypothesis with at least one significant microstructural or macrostructural metric in adjacent resolution pairs in every bundle. We correct for multiple comparisons using a false discovery rate (FDR) estimation [33] (see Supplemental material A with the raw and FDR corrected $p$-values for this experiment).

## 3. Results

3.1 Quantitative Results

Our Wilcoxon Signed-Rank test showed statistical significance ($p \leq 0.05$) between at least one bundle (macrostructural) or one DTI (microstructural) metric in adjacent resolution pairs (pairs of resolutions with a difference of 0.25 in the $z$ dimension). The macrostructural metrics of average length (AL), surface area (SA), and bundle volume (V), show significant effects across different voxel sizes in all bundles. Once we reject the null hypothesis, we up sample the down sampled images to test our hypothesis about the potential to use spatial sampling for harmonization and used the Cohen's $d$ coefficient to measure the effect size between the original resolution images and those with some degree of information loss from the simple down sampling. The coefficients suggested by Cohen to measure effect size are: $d = 0.2$ indicates small effect size; $d = 0.2$ indicates medium effect size, and $d = 0.2$ indicates large effect size [34]. For the sake of this study, we group $d \leq 0.2$ as small effect size; $0.2 < d < 0.8$ as medium effect size; and $d \geq 0.8$ as large effect size (Figure 3).

Our strategy to resample the voxels to 1.25 mm isotropic original resolution and 1 mm isotropic had mixed results based on metric and bundle. We share the results for the right arcuate fasciculus (AF), the right corticospinal tract (CST), and the corpus callosum (CC), as each of these bundles represents a different major white matter connection direction (anterior-posterior, superior-inferior, and right-left). All three bundles exhibited a range of significance in the $p$-values of our selected metrics (see supplemental material A – specifically in MD and average length), but TractSeg generated these regions in every case despite high anisotropy in many of the voxel sizes. For the microstructural metrics, we see a large effect size across all bundles in both FA and MD. Many of the effect sizes between adjacent down sampled resolutions show smaller effect sizes than those in the reshaped isotropic resolutions. We cannot claim that reshaping the voxels from anisotropic to isotropic produces a more consistent result in the microstructure measurements. In the macrostructure, the resampled resolutions show significant improved consistency in the CST and moderate improved consistency in the AF, but the interpolation in the anisotropic voxels appear to be too significant in the CC to generate consistent results with a basic up sampling strategy, especially in the surface area and volume measurements.

3.2 Qualitative Results

To understand the qualitative changes to microstructure due to resampling, we show the fractional anisotropy maps of the same image at 1 mm isotropic, native 1.25 mm isotropic resolution, $1.25 \: x \: 1.25 \: x \: 2.75$ mm$^3$ (Figure 4). We see that the Gaussian smoothing that occurs between the original resolution and the anisotropic sampling causes blurring in the FA maps, leading to less contrast and detail between anatomical structures and less consistent quantitative metrics. When up sampled, the *mrgrid* default interpolation does not restore the original contrast of the image. The lack of restoration of the original contrast leads to inconsistent results for voxel-based operations like those performed to generate quantitative maps.

Similarly, we display three representative bundles for all major brain connection directions: anterior-posterior (arcuate fasciculus), superior-inferior (corticospinal tracts), and left-right (corpus callossum). We show each of the sets of bundles overlaid on the common T1-weighted image in MI-Brain [35]. We find that tractography generated on smaller, isotropic voxels in the AF and the CST show more localized, concentrated streamlines in a single area, but the CC did not overcome the anisotropic voxel biases as effectively.

4. Discussion

Anisotropic sampling introduces an especially interesting problem in the macrostructural metrics we selected (FA and MD) that do not appear to improve when resampled to a higher resolution. The effect of the Gaussian smoothing on the anisotropic voxels appears to introduce a bias too invasive to the integrity of the image for consistent correction in the voxel-wise operations necessary to calculate microstructural metrics.

Though the resampling strategy may not correct biases introduced by anisotropic voxels and Gaussian smoothing in an image, it is important to note that tractography approaches, such as TractSeg, rely on FODs instead of a tensor model [12], [30]. Though microstructural metrics may be sensitive to down sampling in quantitative mapping tasks, the FODs are able to bypass this issue and generate more repeatable tractography.

4.1 Limitations

Though we fit a tensor model to generate quantitative brain maps based on FA and MD, we do not attempt to generate tractography from these metrics. We acknowledge that tractography based on FODs is state-of-the-art and generated our tractography through the TractSeg algorithm [12], [30], [31]. We did not test if bundles generated by DTI instead of FODs have the same repeatability between low quality, resampled data and high-quality data.

We selected the specific bundles to analyze based on white matter tracts that would reasonably allow us to analyze the tractography from all three major directions (superior-inferior, right-left, and anterior-posterior) and have enough ground truth anatomical surface area for TractSeg to reliably generate, even with high anisotropy. TractSeg is not recommended for smaller anatomical regions with highly anisotropic voxels [30], [31]. It is possible that our selected bundle reproducibility may vary based on method or change if we generated whole-brain tractography and then segmented out individual regions of interest.

Our methods involve the change in dimension of voxel resolution in the *z* plane, elongating the voxels in a superior-inferior direction. We did not attempt to change the voxel dimensions to induce biases in microstructure and macrostructure in the *x* or *y* directions. We did not attempt to change voxel resolution to have *x* and *y* dimensions different from the acquisition resolution. Results may vary based on resampling directions and the bundles impacted by resampling choices, as individual bundles may be sensitive to different anisotropic spatial sampling.

Our selection of HCP-YA data limits our study to a healthy young adult population, and that the anatomy of subjects of different ages or disease statuses may have various levels of sensitivity to the effect of anisotropic voxels discussed in this work. We attempted only to use *mrgrid*'s default interpolation method, cubic interpolation [26]. We did not try more complex interpolation methods or any super resolution algorithms to recover the true anatomy of the smoothed data.

5. Conclusion

In this work, we have identify microstructural or macrostructural metrics for bundles generated by anisotropic voxels that differ in 0.25 mm steps in the *z* dimension. We learned that quantitative maps of DTI (FA and MD) undergo large effect size changes at every step when images are down sampled in one dimension. Even though we selected three different white matter tract bundles, each showed sensitivity to this kind of down sampling on both the FA and MD map. We learned that the image operations performed to generate the low-quality data are too great to overcome the biases introduced by anisotropic voxels with simple cubic interpolation to reshape the voxels to be high dimensional and isotropic. We did, however, find that up sampling the voxels from anisotropic to isotropic, especially to 1 $mm^3$ isotropic, helped to generate more repeatable tractography in some bundles. Since we mitigated the effect of anisotropic voxel biases on two of our three test bundles, we recommend reshaping voxels to 1 mm isotropic before generating tractography to avoid biases introduced by anisotropy.

Acknowledgements

The Vanderbilt Institute for Clinical and Translational Research (VICTR) is funded by the National Center for Advancing Translational Sciences (NCATS) Clinical Translational Science Award (CTSA) Program, Award Number 5UL1TR002243-03. The content is solely the responsibility of the authors and does not necessarily represent the official views of the NIH. This work was conducted in part using the resources of the Advanced Computing Center for Research and Education at Vanderbilt University, Nashville, TN. ADSP U24AG074855, NIH 1R01EB017230 Uncertainty in Diffusion MRI. This work was supported by Integrated Training in Engineering and Diabetes, grant number T32 DK101003 and National Cancer Institute (NCI), Grant/Award Number: R01 CA253923 and P50HD103537 VKC. The content is solely the responsibility of the authors and does not necessarily represent the official views of the NIH.

We used generative artificial intelligence (AI) to create code segments based on task descriptions, as well as to debug, edit, and autocomplete code. Additionally, generative AI technologies have been employed to assist in structuring sentences and performing grammatical checks. The conceptualization, ideation, and all prompts provided to the AI originated entirely from the authors' creative and intellectual efforts. We take accountability for the review of all content generated by AI in this work.

Figures

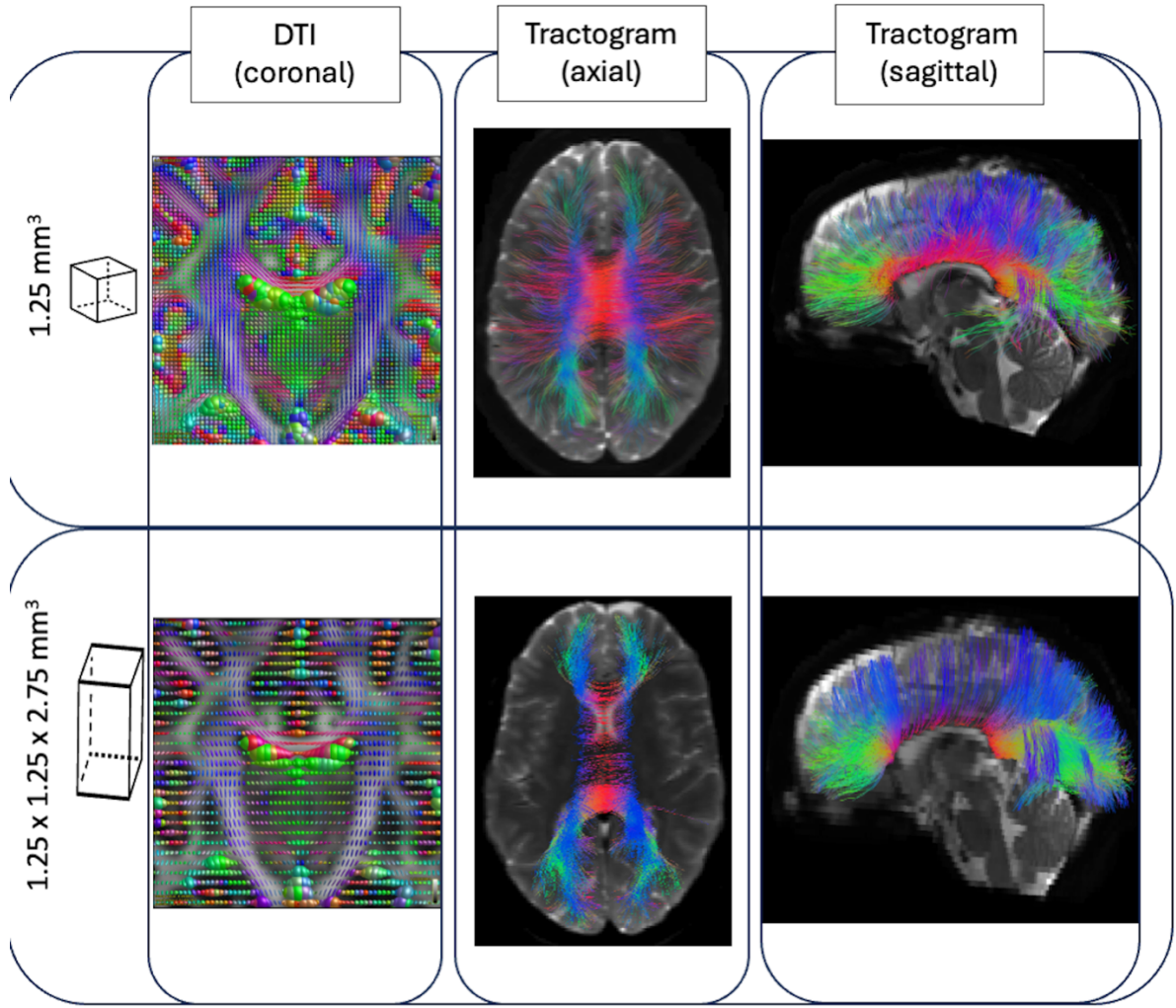

Figure 1: We illustrate the range of voxels used for this experiment with tensor and tractogram representation. We see a bias in the tensor model toward the superior-inferior direction in the anisotropic voxels when compared to the isotropic sampling. The tractogram's representation of the corpus callosum dramatically changes based on spatial sampling; the highly anisotropic voxels influence the tracking behavior to generate superior-inferior streamlines when the corpus callosum's anatomy includes more right-left white matter.

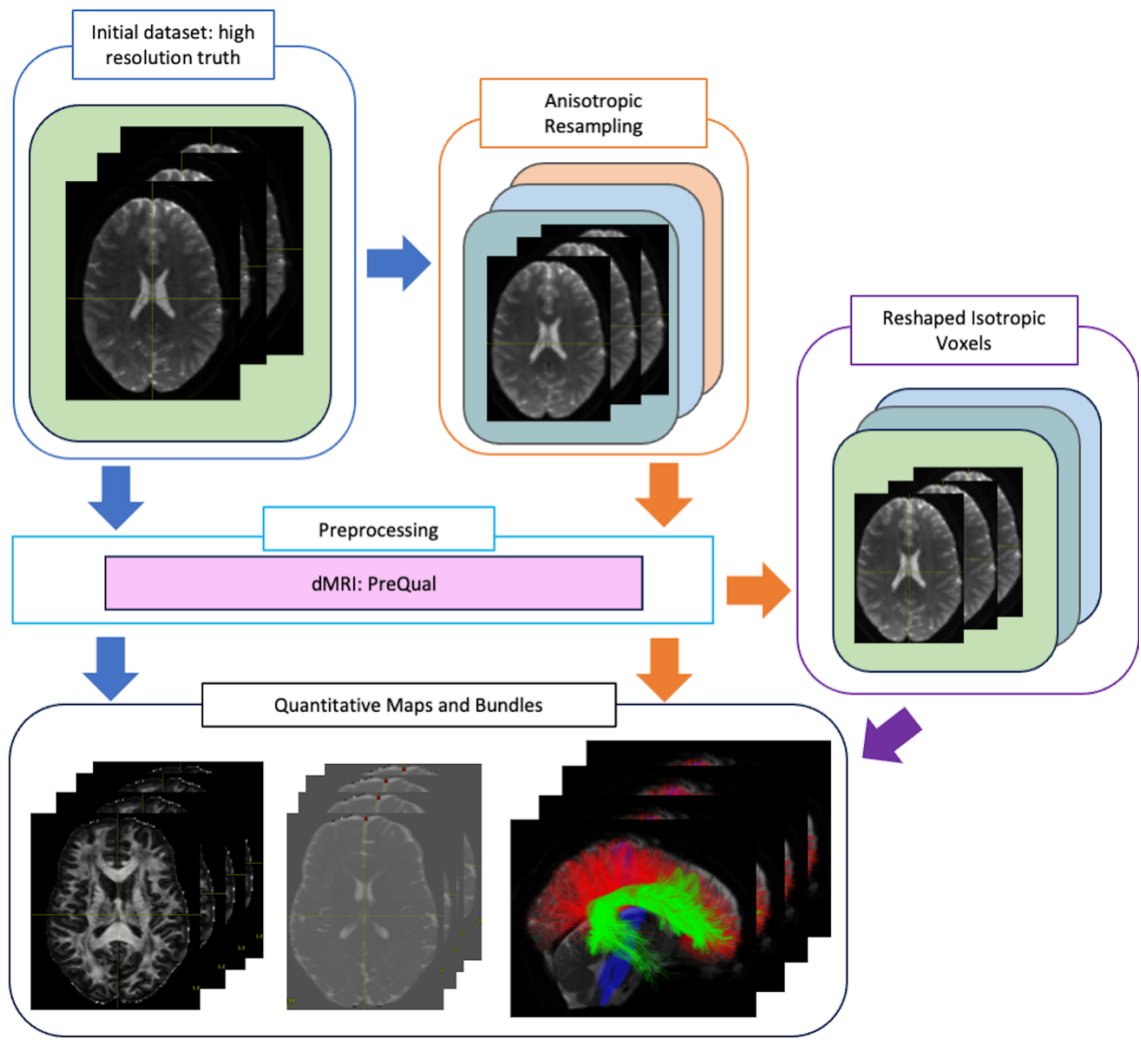

Figure 2: We compare a high-resolution ground truth image to a simulation of low-quality data by down sampling to an anisotropic resolution and then up Figure 3: We show the effect size between resolutions in five methods with three representative tracts. We observe that FA and MD do not become more repeatable across resolutions when resampled to an isotropic voxel, but the tractography metrics show more repeatability in the CST and AF.
sampling to isotropic.

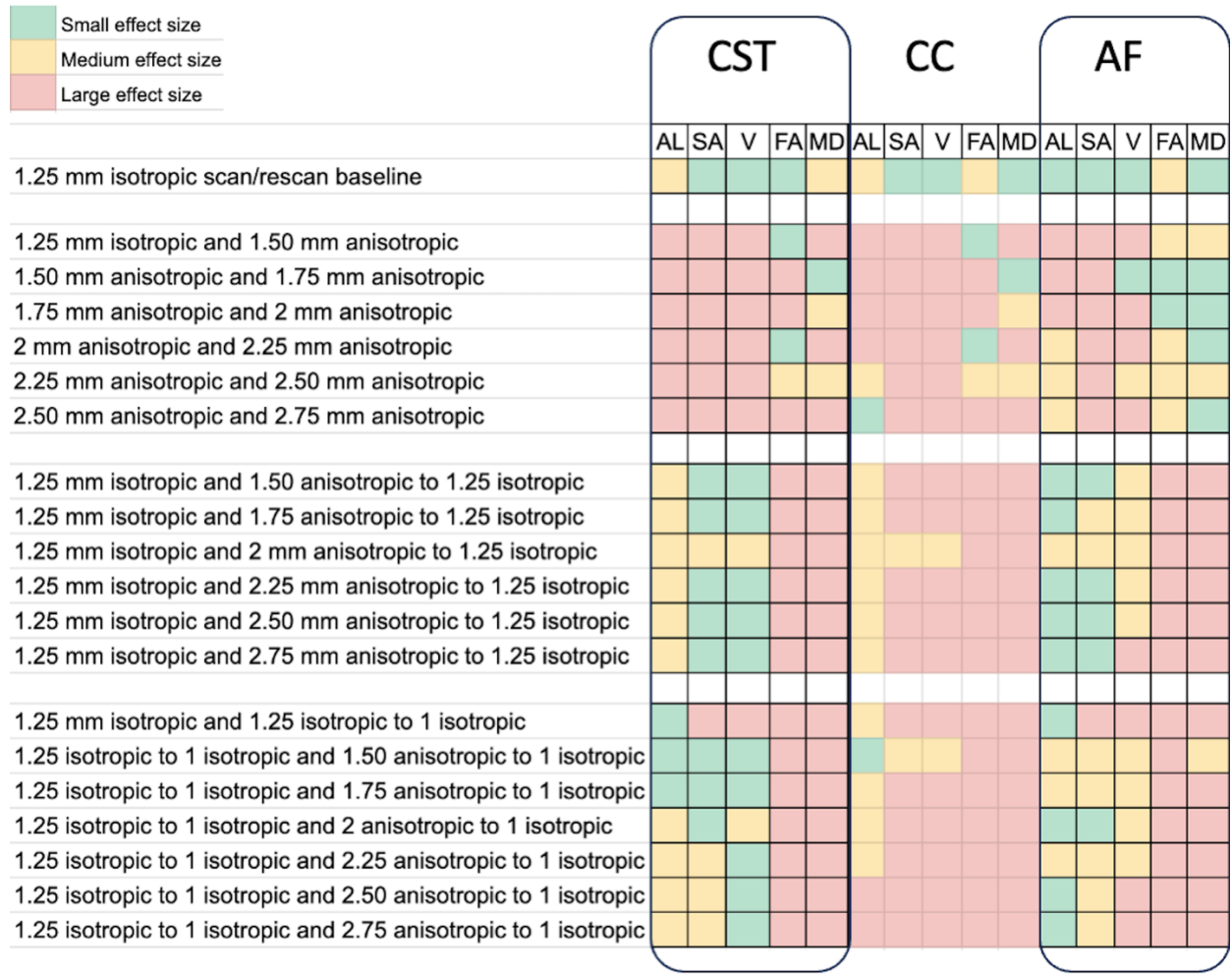

Figure 3: We show the effect size between resolutions in five methods with three representative tracts. We observe that FA and MD do not become more repeatable across resolutions when resampled to an isotropic voxel, but the tractography metrics show more repeatability in the CST and AF.

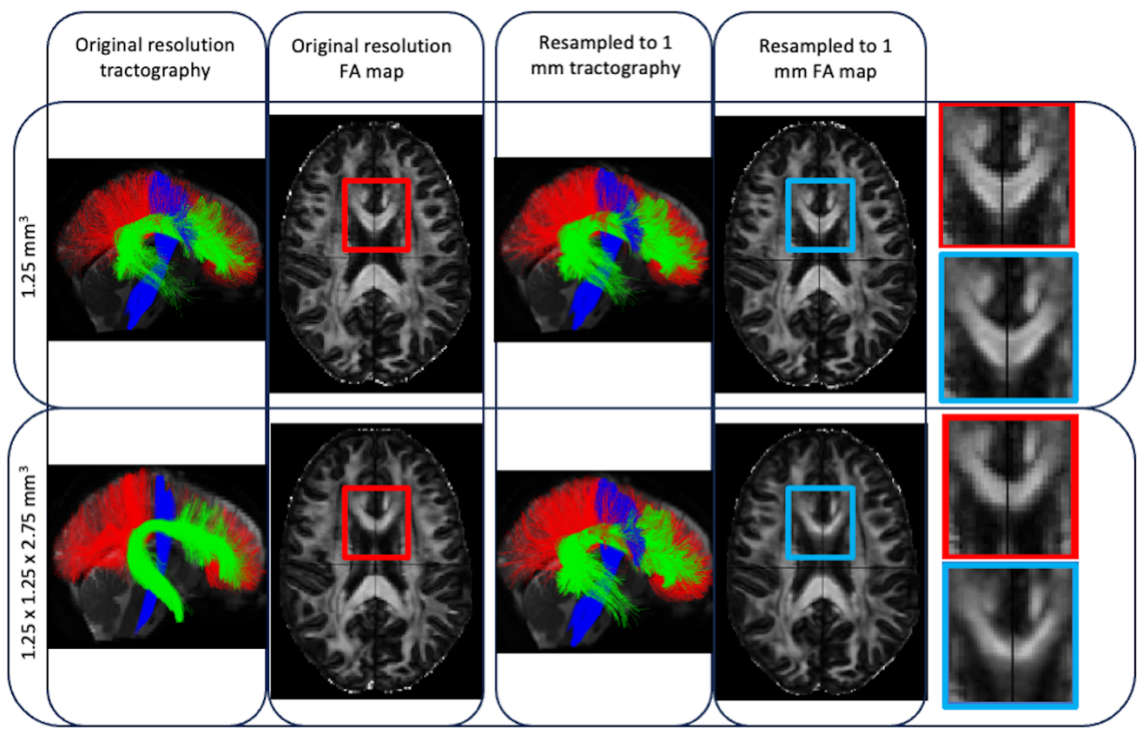

Figure 4: When we compare the tractography between the original isotropic and highly anisotropic voxels, we see that anisotropy leads to thicker, less precise bundles that reshaping the voxels to 1 mm isotropic corrects. In the FA maps, however, we observe a smoothing in our region of interest in the image with high anisotropy that cannot be corrected by resampling.